\documentclass[fleqn,usenatbib]{mnras}

\usepackage{newtxtext,newtxmath}

\usepackage[T1]{fontenc}
\usepackage{ae,aecompl}

\usepackage{graphicx}	
\usepackage{amsmath}	
\usepackage{amssymb}	
\usepackage{booktabs}   

\def\dbdt{${\rm d} b / {\rm d} t$}

\title[Kepler-13Ab with TESS]{The clockwork is moving on - a combined analysis of TESS and Kepler measurements of Kepler-13Ab}

\author[Gy. M. Szab\'o et al.]{
Gy.\ M.\ Szab\'o,$^{1,2}$\thanks{E-mail: szgy@gothard.hu}
T. Pribulla,$^{1,2,3}$
A. P\'al,$^{4}$
A. B\'odi,$^{4,5}$
L. L. Kiss,$^{2,4,6}$
A.\ Derekas$^{1,2,4}$
\\
$^{1}$ ELTE E\"otv\"os Lor\'and University, Gothard Astrophysical Observatory, Szombathely, Hungary\\
$^{2}$ MTA-ELTE Exoplanet Research Group, 9700 Szombathely, Szent Imre h. u. 112, Hungary\\
$^{3}$ Astronomical Institute of the Slovak Academy of Sciences, 059~60 Tatransk\'a Lomnica, Slovakia\\
$^{4}$ Konkoly Observatory, Research Centre for Astronomy and Earth Sciences, H-1121 Budapest, Konkoly Thege Mikl\'os \'ut 15-17, Hungary\\
$^{5}$ MTA CSFK Lend\"ulet Near-Field Cosmology Research Group, Konkoly Thege Mikl\'os \'ut 15-17, H-1121 Budapest, Hungary\\
$^6$ Sydney Institute for Astronomy, School of Physics A29, University of Sydney, NSW 2006, Australia
}

\date{Accepted XXX. Received YYY; in original form ZZZ}

\pubyear{2019}

\begin{document}
\label{firstpage}
\pagerange{\pageref{firstpage}--\pageref{lastpage}}
\maketitle

\begin{abstract}
{ Kepler-13Ab (KOI-13) is an exoplanet orbiting a rapidly rotating A-type star. The system shows a significant spin-orbit misalignment and a changing transit duration most probably caused by the precession of the orbit}. Here we present a self-consistent analysis of the system combining {\it Kepler} and TESS observations. We model the light curves asssuming a planet transits a rotating oblate star which has a strong surface temperature gradient due to rotation-induced gravity darkening. The transit chord moves slowly as an emergent feature of orbital precession excited by the oblate star with a decline rate in the impact parameter of ${\rm d}b/{\rm d}t = -0.011/{\rm yr}$, and with an { actual value of $b=0.19$ for} the latest TESS measurements. The { changing transit duration that was measured from} {\it Kepler} Q2 and Q17 quarters and the TESS measurements { indicates} a linear drift of the impact parameter. 
The solutions for the stellar spin axis suggest a nearly orthogonal aspect, with inclination { around  $100^\circ$}.
\end{abstract}




\section{Introduction} \label{sec:intro}

Kepler-13Ab is { a wide binary system 
with an orbital period of about 10\,000 years. The apparent separation of the components is, however, only about 1.1\arcsec, thus all space photometric light curves contain the combined light of the two stars. The binary harbors a massive planet with an orbital period of 1.76~days \citep{bor11} at a distance from its host of only 0.034~au. It has already been shown by \citet{sza11} that the planet orbits the brighter component of the visual pair. \citet{howell19} were the latest to confirm this conclusively.}

The host star is a rapidly rotating A-type star \citep[$v \sin{i_\star}\approx70\,{\rm km~s}^{-1}$;][]{sza11,shp14}.
The transit shape is asymmetric \citep{sza11} which is due to spin-orbit misalignment \citep{sza11,bar11}. Since its discovery, Kepler-13Ab has been extensively investigated using different approaches. \cite{bar11}, \cite{mas15} and \cite{her18} used gravity-darkened transit models to characterise the spin-orbit misalignment, while \cite{how17} applied the model described by \cite{esp11} with the approximation that the rotational distortion of the host star's surface corresponds to a Roche surface. 

Transit duration variation was detected by \cite{sza12} and later confirmed by \cite{mas15}. The oblateness (due to rotation) of Kepler-13A causes secular perturbations and, as a consequence,
the planetary orbit precesses. \cite{sza12} revealed the shifting of the transit path towards lower latitudes and predicted secular variations in the light-curve shape, such as its depth and asymmetry. \cite{mas15} fitted the precession model to the time series of $i$, $\lambda$, and $i_{\star}$ obtained with the gravity-darkened model and constrained the stellar quadrupole moment. Their result yielded a smaller value for $J_{2}$ than it was previously determined from the observed transit duration variations (TDVs) and suggested that the difference will be observable in future follow-up observations, thus in the evolution of $\lambda$.

The Transiting Exoplanet Survey Satellite \citep[TESS;][]{2015JATIS...1a4003R},
has offered for the first time those long-awaited follow-up observations with enough precision and time coverage. After its initial southern survey, TESS turned to the north in mid-2019, covering the {\it Kepler} field in S14 and S15. Here we combine the new space photometry of Kepler-13Ab with the archival {\it Kepler} data to update our knowledge of this intriguing exoplanet system.

\section{Observations} 

Here we present our analysis of { Kepler-13Ab} based on the TESS observations. TESS data were obtained both in the S14 and S15 sectors, between JD 2\,458\,683 and 2\,458\,737. In total, 35417 individual short cadence data points were taken and analysed in this study. { The mission-provided publicly available\footnote{https://mast.stsci.edu/} SAP light curves were used.}

{ We used {\tt WOTAN} \citep{wotan}  to de-trend the out-of-transit part of the signal.} Data points in two time intervals were omitted because of increased scatter and prominent instability of systematic noise (JD 8691.71 -- 8692.23 and 8697.32 -- 8698.11). Due to the photometric noise, the data set was folded into a single phase diagram for visualisation purposes, and it was also binned in the phase space by 40 points. Although we illustrate the results with the binned phased light curve, we note that the entire analysis was based on the unbinned time series.

For comparison purposes, {\it Kepler} Q2 and Q17 quarters were also included in the analysis, { and we again used {\tt WOTAN} to de-trend the PDC light curves.} We applied the iterative biweight method with a time window of 0.4 days and performed the same fitting procedure as with the TESS time series.

\section{Methods} 

The light curves were first sent through the quadratic limb darkening transit model analysis \citep[see][]{man02,pal2008}. This results in a fit with four parameters, which are related to the transit timing, duration, depth, and the curvature of the light curve. The fits were calculated with the same code based on the FITSH/lfit program \citep{pal2012} as we used in previous studies \citep[e.g.][]{sza11,sza12}. 

The planetary transits were then modelled using a more complex algorithm including free orientation of the stellar rotational axis, stellar oblateness and gravity darkening due to the rotation, improved 5-parameter limb-darkening law of \citet{cla18}, and the rotational Doppler beaming. 

{ The first step in the synthesis of the transit light curve was rendering of a rotationally-deformed stellar shape defined by the following equipotential:

\begin{equation}
\psi = -\frac{GM}{r} - \frac{1}{2} \Omega^2 r^2 \sin \theta,
\end{equation}

\noindent where $G$ is the gravitational constant, $M$ is the mass of the star, and $r$ is its radius on colatitude $\theta$. The above equation was then parameterised by the ratio of the stellar angular rotation velocity, $\Omega$, to the break-up velocity, $\Omega_{\rm crit}$. Then the 3D shape was projected onto the plane of the sky for a given stellar rotational axis inclination angle, $i_\star$. The projected view was rendered in equidistant pixels. The typical radius of the projected stellar shape appropriate for the satellite photometry accuracy was 200-300 pixels. In the next step, the local gravity and corresponding effective temperature was computed for each pixel. The gravity darkening was modelled using the analytical approach of \citet{esp11} appropriate for radiative stellar envelopes. Improved 5-parameter limb-darkening law with $\mu_{\rm crit}$ was used \citep{cla18}, interpolating for local gravity and temperature for each pixel. Finally, the total out-of-the-transit stellar flux, as seen by the observer, was computed. The black-body approximation was used for a given filter central wavelength. The Doppler beaming correction was taken into account. 

The transit light curve was obtained by subtracting the flux from the part of the star eclipsed at a given orbital phase from the out-of-eclipse flux. The light curve was synthesized for a fixed number of equidistant phases, typically 1800 points. The resulting code uses the following parameters: time of the periastron passage $T_0$ (for circular orbits equal to mid-transit time if $\omega = \pi/2$ is used), orbital period $P$, orbital inclination $i$ at a pre-defined epoch, orbital inclination change rate $di/dt$, ratio of the stellar radius to semi-major axis $R_\star/a$, ratio of planetary and stellar radii $R_p/R_\star$, orbital eccentricity $e$, longitude of the periastron passage $\omega$, stellar projected rotational velocity $v \sin i_\star$, projected orbital-plane rotational axis misalignment angle $\lambda$, inclination of the stellar rotational axis $i_\star$, third light $l_3$ and light-curve normalization factor $l_{\rm norm}$. The break-up velocity, $v_{\rm crit}$, is determined as:

\begin{equation}
v_{\rm crit} = \sqrt{GM/R_E},
\end{equation}

\noindent where $R_E$ is the equatorial radius of the star.\footnote{ Given that we deal with an oblate star, we use $R_\star$ = $R_E$ in further considerations.} Using the observed $v \sin i_\star$ and stellar rotational axis inclination angle $i_\star$ gives $\Omega/\Omega_{\rm crit}$. }

\begin{figure}
    \centering
    \includegraphics[viewport=0 55 270 132,width=\columnwidth]{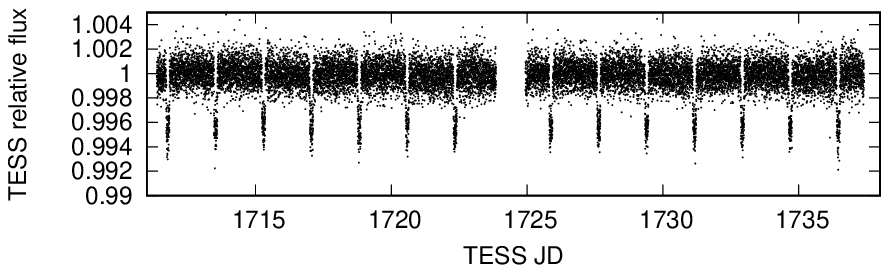}
    \includegraphics[width=\columnwidth]{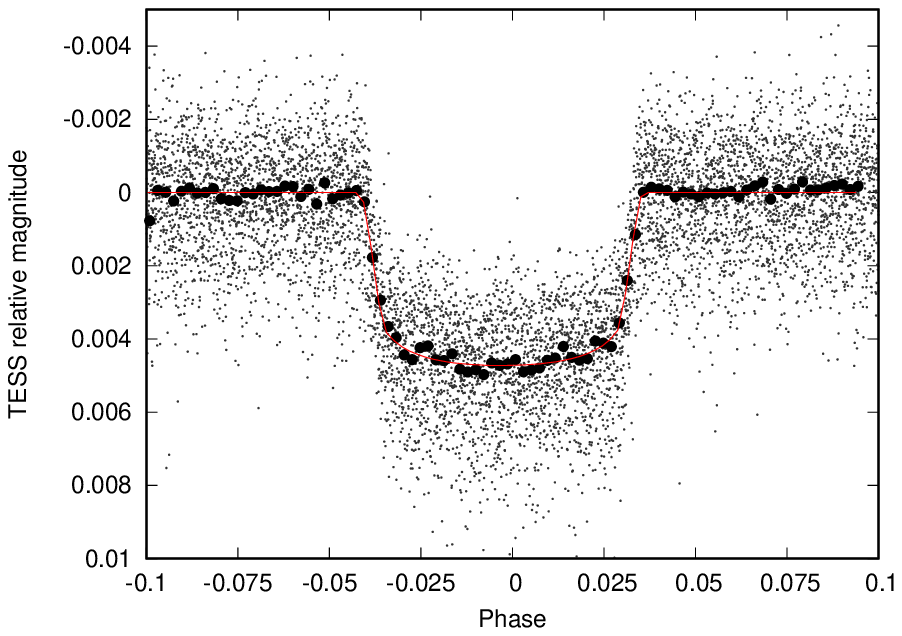}
    \caption{TESS light curve of Kepler-13Ab. { Upper panel: raw S15 light curve, lower panel: phased S14 and S15 observations.} TESS measured A and B components together, hence the transit depth is compressed.}
\end{figure}

\begin{figure}
\includegraphics[viewport=0 30 270 165, width=\columnwidth]{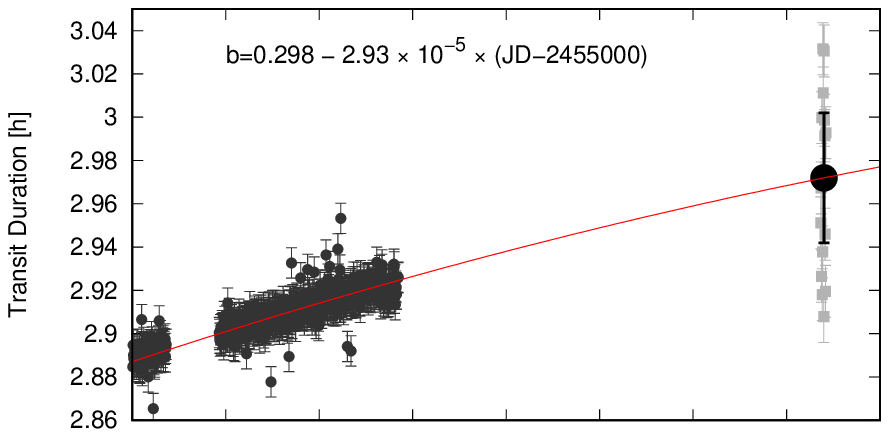}
\includegraphics[viewport=5 20 270 165,width=\columnwidth]{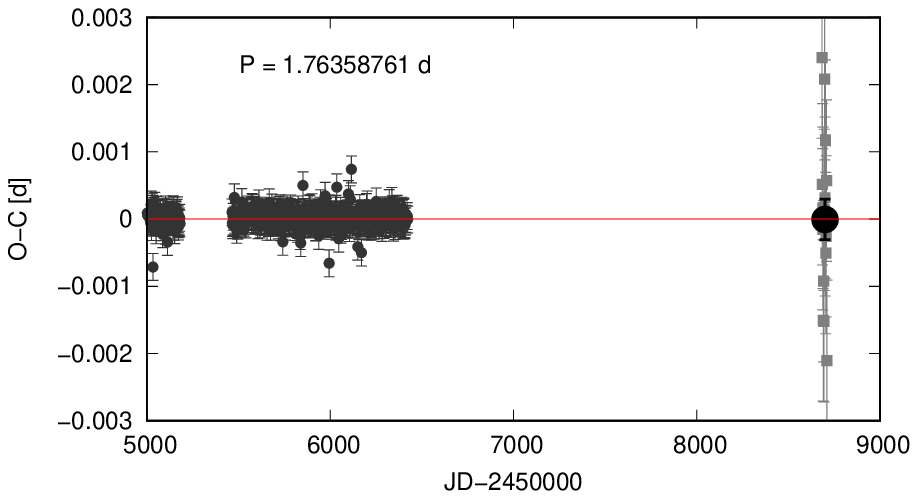}
\caption{Upper panel: Transit duration of Kepler-13Ab. Duration data from {\it Kepler} photometry are plotted individually for all transits in Q2--Q17, TESS data were fitted in the folded transit light curve. Bottom panel: the absence of any transit timing variation (TTV) is very strongly constrained by the two sources of data. \label{impact}}
\end{figure}

{ For Kepler-13Ab we adopted the following assumptions.}
Polar parameters were set to $T_{\rm eff}$ = 8600 K, and log{ $g$} = 4.173 [cgs]
{\bf and another model was also calculated with $T_{\rm eff}$ = 8000 K, to check the parameter correlations with the temperature.}. { A circular orbit was assumed.} Stellar mass and radius{, $M = 1.72\pm0.10$ M$_\odot$ and $ R = 1.71\pm0.10$ R$_\odot$,} determined by \citet{shp14} from Keck spectroscopy,  were used.
{ Using these parameters and neglecting the difference of the mean and equatorial radius, we get $v_{\rm crit}$ = 438 km~s$^{-1}$. For the observed projected rotational velocity $v \sin{i_\star}$ = 76.96 km~s$^{-1}$ \citep{john14}, this implies $\omega_{\rm rot} = \Omega/\Omega_{\rm crit} \geq 0.176$, depending on the { inclination of the }stellar rotational axis. The minimum rotational axis inclination is about 10\degr. The projected rotational velocity was kept fixed adjusting $\omega_{\rm rot}$ according to $i_\star$ in the modelling.} 

The gravity-darkening effect due to the stellar rotation and the spin-orbit misalignment result in the asymmetric transit light curve.  The projected misalignment was reliably determined to be 58.6$\pm$2.0\degr
with Doppler tomography by \citet{john14}. This is significantly inconsistent with the photometric determination of \citet{bar11}, indicating strong parameter correlations when only photometric data are used. Hence, in our modelling we adopted the  robust spectroscopic determination of \citet{john14}.

{ The {\it Kepler} Q2, Q17 and the TESS data were fitted together, assuming a linear trend in the inclination angle, $di/dt$ = const, to draw a consistent picture.} In the fitting procedure, local monochromatic intensities corresponding to a black body at the central effective wavelengths of 596~nm ({\it Kepler}) and 737~nm (TESS) were assumed\footnote{Effective wavelength was determined as the blackbody-flux weighted average over the satellite response function.}. { Third light, resulting from the contribution of visual component B }, was assumed to be 0.91 and 0.93\footnote{Defined here as the ratio of the third object flux and the out-of-eclipse flux of the exoplanet parent star.} { for} {\it Kepler} and TESS { central wavelengths}, respectively \citep{shp14}.

For the Doppler beaming effect the beaming efficiency coefficient $\alpha=1$ was used. The amplitude of the beaming effect across the transit is only about 3~ppm. In the fitting procedure, all individual datapoints were fitted (after omitting the few points influenced by instrumental systematics). { A steepest descent algorithm was used to arrive to the optimum parameters. Multiple starting parameter sets were attempted to ensure that the global minimum of the merit function ($\chi^2$) would be reached. The formal errors of {\it Kepler} and TESS photometry were used. The parameter errors (Table~1) were derived from the covariance matrix and no employed to take into account systematics was done. Presence of some red noise and systematic errors is indicated by reduced $\chi^2_r \sim 1.4$.}

\section{Results} 

In Fig. 1 we plot the S14-S15 TESS observations, and the binned phase diagram. The red line shows the best fitting \citet{man02} model, which now looks satisfying since TESS observations were too noisy to show the asymmetric light-curve shape by eye. Two parameters of this standard fit are of special importance in the case of this system, since the duration of the transit is known to gradually increase. 

The transit duration can be directly related to the $b$ impact parameter, which is the relative distance between the transit chord and the center of the stellar disk, so the central transit has an impact parameter of 0, and the grazing transit can be described by $b=1$. If $b$ varies due to orbital precession, { neglecting a small oblateness of the star}, the transit duration will be proportional to $\sqrt{1-b^2}$. 

In Fig. 2 upper panel, we plot the increasing transit duration. { All transits were fitted individually, and according to the recipe, a symmetric template curve was used  (quite similarly as we show the fits in Fig. 1). Because of the larger scatter, parameters of the TESS transits (in grey points) were averaged (black point with an error bar).} The TESS data perfectly confirm the variations that were discovered in the {\it Kepler} data. The best-fit linear function suggests a \dbdt{} value of $-$0.011~yr$^{-1}$. This is within the error bars of \citet{sza12} and is found at the lower range of the confidence interval. { We emphasize here that \dbdt{}{}, which is a parameter of the red curve, was not at all fitted to the data points we plot here at all, but was derived from our separate analysis of the light curve asymmetry, as we describe later and show in Fig. 3. Still, the red curve beautifully fits the general trend and the fine structures of the black points, which proves the general self-consistency of the analysis in this paper.}

In the lower panel of Fig. 2 the Transit Timing Variation (TTV) is shown. Orbital elements can change due to outer perturbers as well, but in this case a TTV is expected \citep{2012ApJ...750..114F}. We see no TTV in the lower panel, proving that the reason of the transit duration variation is the stellar rotation and the orbital precession, and not an outer perturber.

In Fig. 3 we see a comparison of transit chord for { simultaneous} solution for {\it Kepler} Q2, Q17 and TESS measurements. The best fit values of the fitted parameters are summarised { in Table~\ref{elements}. The local monochromatic flux, given by the local temperature and limb darkening, on the stellar disk is shown, and the green--yellow--red--magenta coloring scheme indicates the increasing flux.} The stellar disk is not perfectly circular, but is a bit oblate because of the rotation.

\begin{table}
\centering
\begin{tabular}{l|cll}
\hline
Parameter      & unit      &   $T^{\rm eff}=8000$K   &  $T^{\rm eff}=8600$K     \\
\hline
 $P$           & [day]     &  1.76358762(3)          &  1.76358760(3)           \\
 $T_0$         & [HJD]     &  55\,101.707249(13)     &  55\,101.707254(12)      \\
 $R_\star/a$   &           &  0.22754(11)            &  0.22880(11)             \\
 $R_p/R_\star$ &           &  0.08606(3)             &  0.08632(3)              \\
 $i_0$         & [deg]     & 86.084(19)             &  85.738(18)              \\
 ${\rm d}i/{\rm d}t$       & [deg/day] 
                        & $3.83(13)~10^{-4}$     &  3.45(12)~10$^{-4}$     \\
 $i_\star$     & [deg]     & 100.5(8)                &  102.5(8)               \\
 \dbdt{}       &[day$^{-1}$] 
                        & $-2.93(10)~10^{-5}$    & $-2.63(9)~10^{-5}$     \\
 \hline
 $\chi^2$      &           &  20838                  &  20924                   \\
 d.o.f.        &           &  13384                  &  13384                   \\
 \hline
\end{tabular}
\caption{Best parameters obtained by the simultaneous modelling of the {\it Kepler} Q2, Q17 and TESS measurements. $i_0$ corresponds to the mid of the {\it Kepler} Q2. Two solutions for different polar temperature of the parent star are shown. Mid-transits times, $T_0$ are given -2\,400\,000.}
\label{elements}
\end{table}

\begin{figure}
    \centering
    \includegraphics[width=\columnwidth]{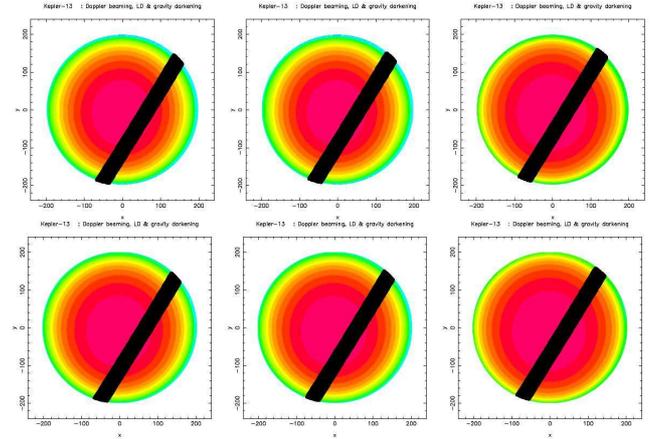}
    \caption{ Simultaneous fit of {\it Kepler} Q2, Q17 and TESS data, accounting for gravity darkening, stellar oblateness and rotational beaming due to stellar rotation. The transit chord in front of the stellar surface is shown in case of $T_{\rm eff}$=8000~K (upper row) and 8600~K (lower row). The columns show {\it Kepler} Q2, Q17 and TESS data, respectively. The green--yellow--red--magenta coloring scheme of the stellar disk reflects the increasing local flux as seen by the observer. The planet crosses the disk from bottom left to top right. The star rotates from the left to the right edge. The southern rotational pole is tilted towards the observer.}
    \label{fig:my_label}
\end{figure}

{ The resulting parameters in Table~\ref{elements} are consistent with previously published results. \citet{bar11} (see their Table~1) got $R_\star/a$ = 0.22369 and 0.22627, $R_p/R_\star$ = 0.084508 and 0.084513, for $M$ = 1.83 and 2.05 M$_\odot$ solutions, respectively. The stellar obliquity determined assuming $\lambda = 58.6\degr$ is consistent with that found by \citet{mas15}. 

{ Models with two polar effective temperatures $T_{\rm eff} = 8000$ K and $T_{\rm eff} = 8600$ K show similar results. The largest differences occur in $R_p/R_\star$ and $i_\star$ and are very probably caused by the different limb and gravity darkening effects.}

{\bf The modeling assumes that we observe only a very short part of the orbital precession cycle, so the change in inclination angle is close to being linear. This is supported by Fig.~\ref{impact} showing the increasing transit duration and the lack of Transit Time Variation.} However, the change of the impact parameter (and the related change in the inclination angle) is detected at high significance even though the shift of the transit path merely exceeds one planet
radius.

\begin{figure}
    \centering
    \includegraphics[viewport=53 156 400 347, width=\columnwidth]{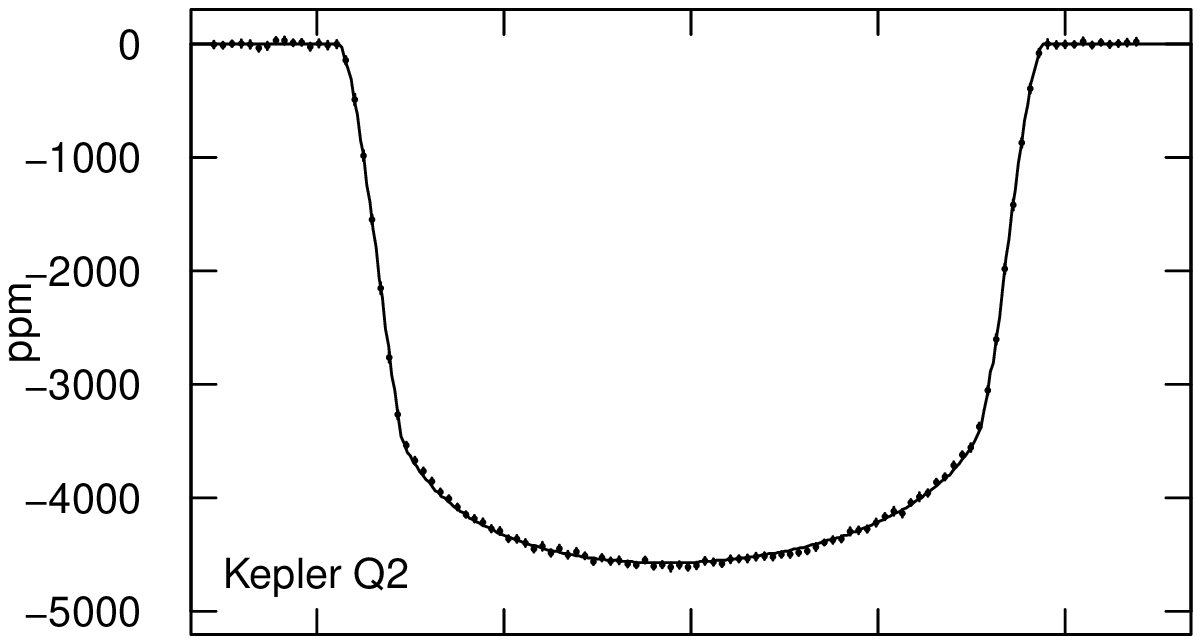}
    \includegraphics[viewport=53 185 400 284, width=\columnwidth]{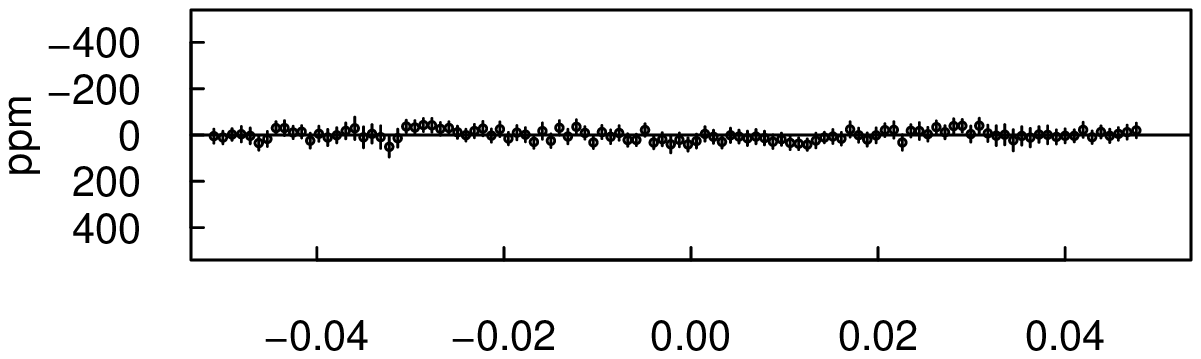}
    \includegraphics[viewport=53 156 400 347, width=\columnwidth]{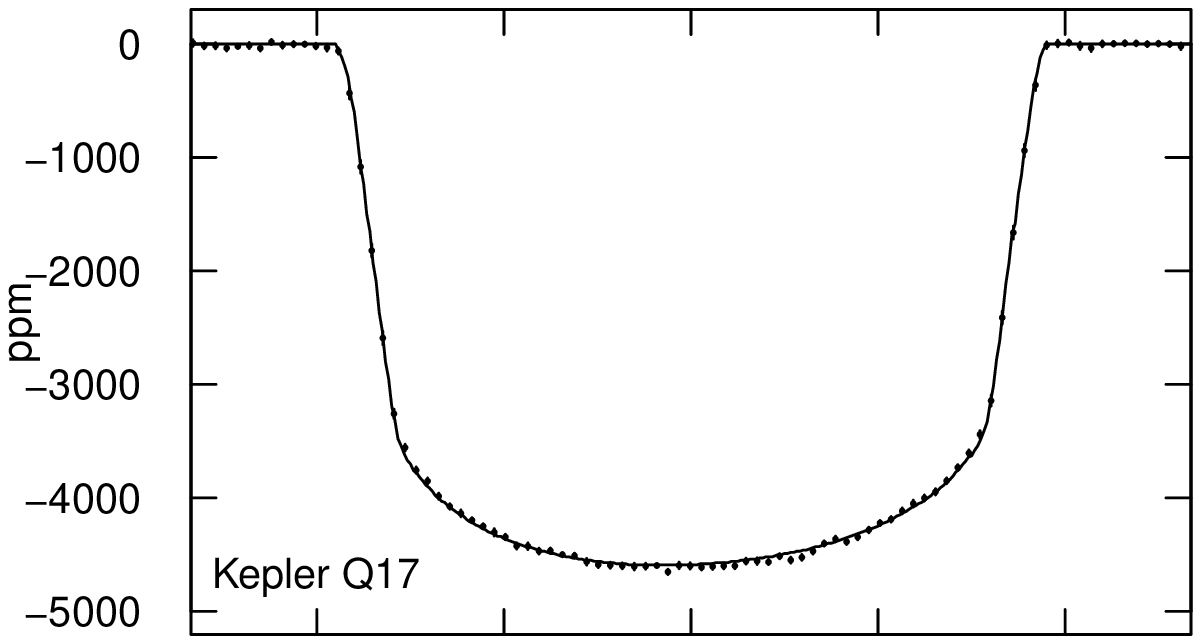}
    \includegraphics[viewport=53 185 400 284, width=\columnwidth]{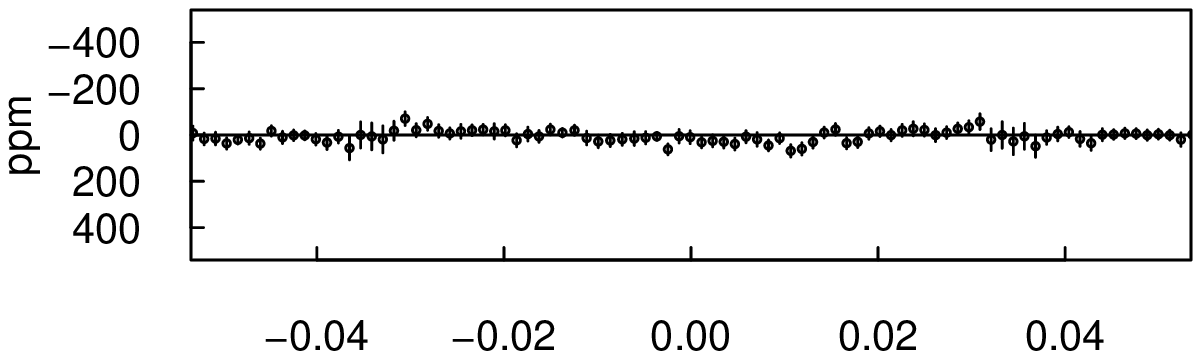}
    \includegraphics[viewport=53 156 400 347, width=\columnwidth]{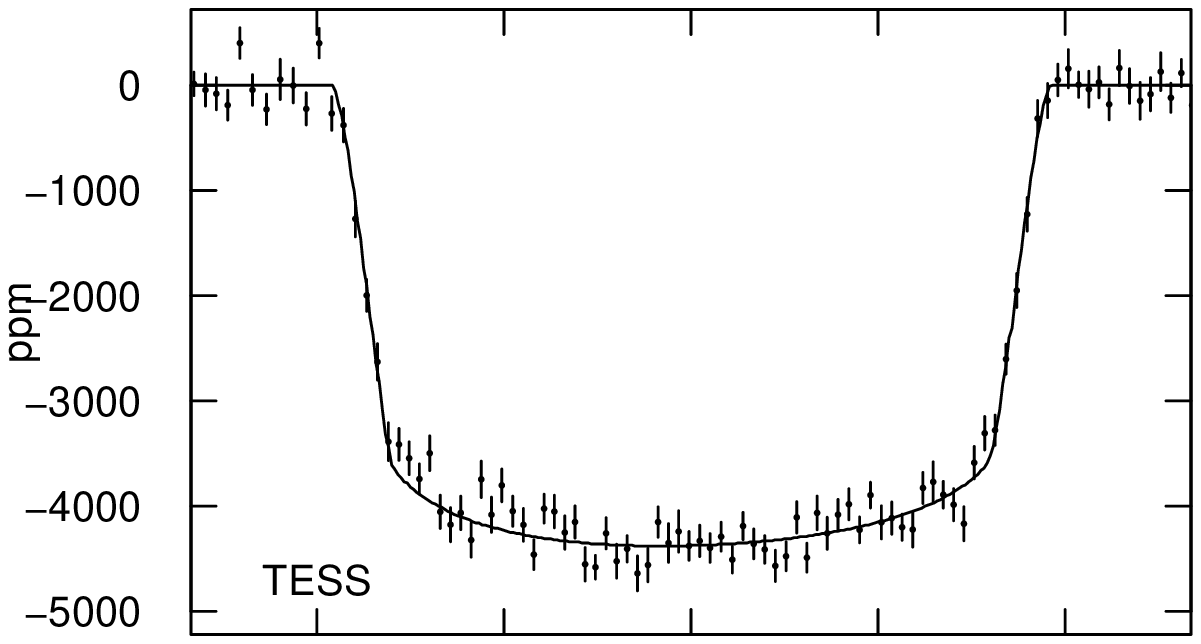}
    \includegraphics[viewport=53 156 400 284, width=\columnwidth]{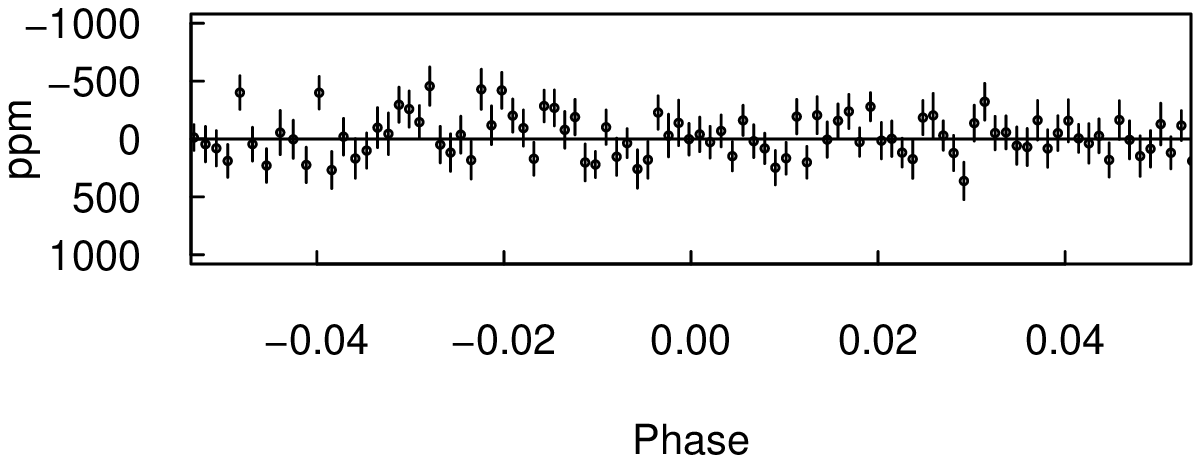}
    \caption{{ The individual fits corresponding to the simultaneous modelling and the binned folded light curves. In the lower panels for each fits, the residuals are shown.}}
    \label{fig:4}
\end{figure}

It is known that the angular momentum vectors of the orbit and that of the star have to precess with the same period in an axial symmetry around their sum, to maintain the conservation of total angular momentum. One can argue \citep[as was expected by e.g.][]{mas15} that during a time base of 10 years, the secular changes in stellar spin will be also revealed. Separate solutions of the {\it Kepler} data showed the stellar inclination to be compatible to each other. The new measurement was expected to precisely unveil the stellar precession, too \citep{mas15}. This is not the case. Because of the larger errors in the TESS measurements, $i_\star$ is consistent with the stellar spin axis solutions for the {\it Kepler} data. Hence, in the simultaneous model we assumed a constant $i_\star$. Thus, new data will be required to get to a firm detection of $i_\star$ variation. As no upcoming space mission is expected to approach the accuracy of the {\it Kepler} photometry, a new Doppler tomography of the Kepler-13Ab transits is necessary to study the nodal precession and to detect the partition of the total angular momentum vector between the orbit and the stellar spin \citep[see the case of WASP-33;][]{john15}.}

\section{Discussion and closing remarks}

{ Kepler-13Ab} is an emblematic system showing many unique aspects of interactions between rotating stars and close-in planets. The shifting transit chord was already recognised in the  {\it Kepler} data, but \dbdt{} could not be reliably fitted directly, and several indirect approaches were applied to estimate its value \citep{sza12,mas15}. Now, a self-consistent solution of the {\it Kepler} and TESS data together has shown directly how the transit chord evolves. This solution is also perfectly compatible with the transit durations observed by {\it Kepler} and TESS. { The red curve in Fig. 2 is actually the evolution of $b$ as derived from the asymmetric models shown in Fig. 4, which are shown in comparison to the transit duration data, as derived from simple symmetric models (similarly to the curve in Fig. 1). So the points and curve in Fig. 2 are not fitted to each other, but show results from two different approaches. Since in Fig. 2, the curve follows the distributions of the points perfectly, it proves the general consistency of our analysis.}

The above fits can well reproduce the motion of the planet and the shift of its orbit, but do not show well the precession of the stellar spin. 

In the fits we present here we assumed a stable $\lambda$, simply because this is a very degenerated variable, and if we allow $\lambda$ to vary, the fits do not tend to converge well. We know that $\lambda$ is not a free parameter, it is fully constrained by the stellar and orbital inclinations via the conservation of angular momentum, but we do not know how exactly.

On one hand, observation of a varying $\lambda$ promises that the partition of the total angular momentum vector between the orbit and the stellar spin can be decomposed from the long-term observation of the dynamics. On the other hand, allowing $\lambda$ to vary during the modeling can introduce quite unexpected instabilities to the fitting procedure. 

The internal structure of the spin-orbit precession still remains unconstrained by the TESS data, and is still waiting for further missions. 

\section*{Acknowledgements}

{ The authors thank to an anonymous referee for his/her valuable comments which helped to improve the article.}
This project has been supported by the Hungarian NKFI Grant  K-119517 and the GINOP 2.3.2-15-2016-00003 of the Hungarian National Research, Development and Innovation Office, and the Lend\"ulet LP2018-7/2019 grants of the Hungarian Academy of Sciences, and  by the Slovak Research and Development Agency
under the contract No. APVV-15-0458. AD was supported by the \'UNKP-19-4 New National Excellence Program of the Ministry of Human Capacities and the J\'anos Bolyai Research Scholarship of the Hungarian Academy of Sciences. AD and GyMSz would like to thank the City of Szombathely for support under Agreement No. 67.177-21/2016. { This paper includes data collected with the TESS and {\it Kepler} missions, obtained from the MAST data archive at the Space Telescope Science Institute (STScI). Funding for the TESS mission is provided by the NASA Explorer Program. Funding for the {\it Kepler} mission is provided by the NASA Science Mission Directorate. STScI is operated by the Association of Universities for Research in Astronomy, Inc., under NASA contract NAS 5–26555.}







\bsp	
\label{lastpage}
\end{document}